\documentclass[pre,twocolumn]{revtex4}%
\usepackage{amssymb}
\usepackage{graphicx}
\usepackage{dcolumn}
\usepackage{bm}
\usepackage{amsmath}
\usepackage{amsfonts}%
\usepackage{eepic}
\input epsf
\newcommand{\ave}[1]{\langle #1\rangle}

\begin{document}
\bibliographystyle{apsrev}

%\title{A random graph approach to distributed load-balancing.}
\title{A Statistical Mechanical Load Balancer for the Web}

\author{Jesse S. A. Bridgewater}
\email{jsab@pobox.com}
\affiliation{University of California, Los Angeles\\
Department of Electrical Engineering }
\homepage{http://www.pobox.com/~jsab}
\author{P. Oscar Boykin}
\email{boykin@ece.ufl.edu}
\affiliation{University of Florida\\
Department of Electrical and Computer Engineering }
\author{Vwani P. Roychowdhury}
\email{vwani@ee.ucla.edu}
\affiliation{University of California, Los Angeles\\
Department of Electrical Engineering }

\begin{abstract}
The maximum entropy principle from statistical mechanics states that a closed 
system attains an equilibrium distribution that maximizes its entropy. We 
first show that for graphs with fixed number of edges one can define a 
stochastic edge dynamic that can serve as an effective thermalization scheme, 
and hence, the underlying graphs are expected to attain their maximum-entropy 
states, which turn out to be Erd\"{o}s-R\'{e}nyi (ER) random graphs. We next 
show that (i) a rate-equation based analysis of node degree distribution does 
indeed confirm the maximum-entropy principle, and (ii) the edge dynamic can be 
effectively implemented using short random walks on the underlying graphs, 
leading to a local algorithm for the generation of ER random graphs. The 
resulting statistical mechanical system can be adapted to provide a 
distributed and local (i.e., without any centralized monitoring) mechanism for 
load balancing, which can have a significant impact in increasing the 
efficiency and utilization of both the Internet 
(e.g., efficient web mirroring), and large-scale computing infrastructure 
(e.g., cluster and grid computing). 
\end{abstract}

\maketitle

\section{Introduction and Motivation}

In the past several years there has been significant progress in the field of
complex networks due to an infusion of ideas from statistical mechanics.
Some of the branches of statistical mechanics that have informed the
study of complex networks include percolation theory 
\cite{newman-percolation, newman-site-bond, newman-robust-percolation, dorogovtsev-percolation},
renormalization group methods \cite{newman-renormalization},
Bose-Einstein
condensation \cite{albert-statistical,barabasi-bose} and partition 
function methods for the study of equilibrium uncorrelated 
\cite{dorogovtsev-statmech, burda2003, burda2001} and correlated \cite{bergcorrelated} random
networks.

These analytical tools  have led to the formulation of a number of protocols
or stochastic dynamics for complex networks that result in predictable
macroscopic properties. For example, preferential attachment based dynamics,
provide both models for how existing networks might have evolved \cite{
newman-social,  newman-growing-social, sexual-preference, lehmann-citation},
as well as how one might engineer ad hoc systems such that the resulting
networks have desired global properties such as power-law degree distributions
and tolerance to both attacks and failures \cite{nima-del-comp,albert-attack}.
Dynamics on such networks, such as percolation and random walks, have led to
efficient algorithms for searching in PL random networks and peer-to-peer
systems\cite{percolation-search}.

In this paper, we build on such studies and introduce a network dynamical 
system such that the steady state degree distribution will be tightly 
clustered around the average value. The well known ER graphs have such a 
tight clustering property: the probability of deviating from the mean 
decreases exponentially with the deviation distance. In order to design such 
a dynamical system, we model it after a physical system with say a fixed 
energy, and make use of the maximum entropy principle. That is, we 
introduce an edge dynamic for  networks with fixed average number of edges 
and fixed number of nodes, and show that the dynamic can never decrease the 
entropy of the system.  Thus, our edge dynamic can be viewed as an effective 
thermalization process, and the fixed average number of edges can be 
considered to be a physical system with constant energy in a statistical 
mechanical sense. The maximum-entropy degree distribution for a network with 
fixed number of edges corresponds to that of ER random graphs with 
a binomial distribution. Thus, the maximum entropy principle dictates that 
the network system would tend towards an ER random graph. 

We provide further analytical results, based on a rate-equation approach,  
which show that the steady state distributions do indeed correspond to ER 
graphs. We also show how {\em local protocols using short random walks} can 
effectively emulate the original global edge dynamics, thus providing a local and 
distributed stochastic algorithm for a network to self-organize itself into 
ER random graphs. We provide extensive simulations of our local dynamics and 
show that the convergence to ER random graphs is robust even if the 
protocols are modified in different ways to suit practical implementation 
requirements.  

We find that the dynamical system studied here can provide an {\em effective 
load balancing paradigm for the distributed resources accessible on the 
Internet}. As the popularity of the Internet has increased, so too has the 
need for highly scalable web server software. Every major web site uses 
mirrors to, among other things, balance the request load over multiple 
servers.  This service is currently provided by companies such as Akamai which 
maintain proprietary overlay networks with tens of thousands of nodes and 
which routinely handle double-digit percentages of total Internet traffic.
An overlay network is simply a virtual network built on top of an existing
network. Overlay networks often add new features not found in the underlying
network or make certain operations more convenient. The most famous example of
an overlay network is the Internet. The Internet consists of computers and
routers which are connected by different physical links (ethernet, ATM,
phone-line, wireless ,etc.), however the Internet Protocol creates a virtual
IP network that allows the networked computers to be addressed without knowledge 
of the physical transport layer; providing a higher level of abstraction is
often a goal of overlay networks. The overlay network that we propose can be 
built directly on of any of the physical transport layers mentioned above or it 
can use the Internet as the underlying network.
Thus an overlay network need not consist of physical links between nodes.  In an 
addressable network such as the Internet, the edges may simply be a table of 
addresses that each node maintains to represent its neighbors in the overlay. 
Of course in networks that are not globally addressable, the overlay 
edges will be actual open links that are continuously maintained over the 
physical network or alternatively each node can maintain a routing table that
gives complete route information on how to reach its neighbors.

A system using the techniques proposed here can provide effective load 
balancing for Open Source software projects and other organizations seeking 
non-commercial mirroring solutions. Good examples of such projects include 
the Linux kernel\cite{linux} and Debian GNU/Linux \cite{debian}. Each 
project has hundreds of mirrors that are chosen arbitrarily by users and 
thus the demand variance of the mirrors can be quite large.  On the other 
hand, if each of the mirrors can automatically redirect traffic to less 
loaded mirrors the users would have a more reliable service, and the servers 
would have a more predictable load. 
%This framework is also appropriate to 
%highly dynamic peer-to-peer systems such as BitTorrent\cite{bittorrent}, 
%which allow users to share their bandwidth to distribute large data files.

To state the problem as simply as possible, we consider a system of $N$ 
comparable-capacity web servers all of which mirror the same set of contents. 
For efficient usage of these resources, one would want to distribute the 
download requests as evenly as possible, so that no server is significantly 
more loaded than others. The question is, can one achieve such a load 
balancing task, without a centralized server or some equivalent mechanism to 
monitor the global state of the network. Numerous load-balancing applications 
for web servers \cite{jw2001webfarm, vc2002web, ma2002web} based on global 
monitoring have been proposed and several open source projects have formed 
to provide capacity and geography load-balancing \cite{supersparrow, ultramonkey, lvs}.  

We present a scheme fundamentally different from  those proposed in the 
literature: instead of monitoring servers and their availability via a 
static network, {\em we create a dynamic overlay network that provides both 
a measure of instantaneous load distribution, and dynamics for job 
allocation and resource update}. The way we adapt our dynamical network 
system to the task of load balancing is as follows. First, a node's 
in-degree is made to correspond to the unused capacity or instantaneous 
estimate of the free resources of a node. Second, the edge dynamics in our 
system are used to perform  the job allocation and resource updating tasks 
for the load balancing process. That is, when a new job arrives the node 
receiving the job chooses, via a random walk (as prescribed by our edge 
insertion dynamic), the node which is going to execute the job.  The target 
node on receiving the job removes one of its incoming edges to reflect the 
reduced availability of its resources.  Similarly, when a node/server 
completes a job, then to reflect its state of being more ready to receive a 
job, it adds an incoming edge to itself (again via a random walk, as 
prescribed by our edge insertion dynamic) to increase its in-degree. 
In steady state, the rate at which jobs arrive would equal the rate at which 
jobs are completed, and hence the underlying network has a fixed average 
number of edges.

Thus, a dynamic overlay network, connecting all the servers, emerges. 
The state of this network (in particular, as indexed by the in-degree 
distribution of the nodes) represents the instantaneous distribution of 
load over all the servers. The job assignment and the resource update steps, 
performed according to the edge deletion and insertion steps in our network 
dynamics, {\em guarantees that the distribution of load will be fair across 
all the servers in the network}: the underlying graph will be close to an ER 
random graph with a binomial degree distribution. 

We have made several simplifying assumptions here so as to show a 
proof of concept for this scheme. For example, we are implicitly assuming 
that the servers have comparable capacities and that any job can be assigned 
to any of the servers.  A complete treatment of how to adapt our approach to 
address these practical issues, is beyond the scope of this paper, and will be
treated in future work. We provide general guidelines how our scheme can be 
generalized in the discussion section of the paper. Nevertheless, {\em the scheme 
proposed here can be implemented as it is}, without any major modifications, 
and {\em increase the efficiency and  utilization of tasks such as web mirroring 
on the Internet}.

This paper has the following structure.  Section \ref{sec:maximal} describes 
our network dynamical system and shows why we should expect it to lead to an 
ER random graph from the perspective of the maximum entropy principle.  
Section \ref{sec:rate_eq} gives a steady state solution for the degree 
distribution of the nodes and thus provides an analytical argument as to why 
the system will converge to ER random graphs. In section 
\ref{sec:walks_simulations}, we show how the dynamics can be simulated via 
random walks on the underlying graphs and with as little global information 
as possible. We also present extensive simulation results demonstrating the 
robustness of the system. Finally in section \ref{sec:load_balancing}, we 
discuss how a load-balancing problem can be efficiently mapped to our 
statistical mechanical system and present more simulation results to show 
the efficacy of our approach.   
\section{Maximum Entropy Principle}
\label{sec:maximal}
It is well known that for a fixed expected number of edges, the maximum 
entropy graph is the ER graph with a binomial degree distribution.  We can 
see this by the method of Lagrange multipliers.
Suppose $p_i$ is the probability that a node has $i$ incoming edges.  The
expected number of edges in the graph is $E = N\sum i p_i$.  Putting into 
place a constraint on the expected number of edges and the normalization of 
$p_i$, we get the following Lagrangian:
\begin{eqnarray*}
\mathcal{L}&=&-N\sum_i p_i \log p_i \\
& & \mbox + N\sum_i p_i \log{N-1 \choose i}\\
& & \mbox{}+ \alpha N \sum_i i p_i + \beta N \sum_i p_i\\
\frac{1}{N}\frac{\partial\mathcal{L}}{\partial p_i}&=& -(1+\log p_i)
+ \log{N-1 \choose i}\\
& &\mbox{} + \alpha i + \beta\\
\end{eqnarray*}
Setting $\frac{\partial\mathcal{L}}{\partial p_i}=0$, and using the fact that
$\sum_i p_i = 1$ and $\sum_i i p_i = E/N$, we get:
\begin{eqnarray}
p_i&=&{N-1 \choose i}q^i(1-q)^{N-1-i}\\
q&=&\frac{E}{N(N-1)} \nonumber
\label{eq_max_ent_dist}
\end{eqnarray}
Recall that an equivalent description of a directed ER graph is as follows: 
for each node there are $(N-1)$ possible incoming edges, each of which is 
selected independently and with probability $q$ (as defined above). 

Next we consider a directed graph with $N$ nodes and $E$ edges.
At each step, a randomly selected existing edge is deleted.
Additionally, a randomly selected absent edge is inserted.
This dynamical system then can be considered as 
a statistical mechanical system, and we may ask if the edge dynamic is an 
effective thermalizing scheme or not; that means, we have to show that the 
entropy of the system after every step never decreases.

We use standard information theory\cite{CoverThomas} notations for entropy:
\[
H(G) = -\sum_g Prob(G = g) \log Prob(G = g) \ .
\]
The entropy of a graph is exactly the average number of bits it requires
to describe its configuration, or equivalently, the $\log$ of the number of
states it is likely to occupy.
Consider the example of an ER random graph, $G_{ER}$, with $E$ directed edges 
present of $E_m = N(N-1)$ possible edges and $N=|V|$ nodes. There are
$|G_{ER}|$ such graphs and each is equally likely since each possible edge
exists in the graph with the same probability.
\begin{equation}
|G_{ER}| = { E_m \choose E}
\end{equation}

Thus the entropy of an ER graph is:
\begin{eqnarray}
H(G_{ER}) &=& -\sum_{g\in G_{ER}} \frac{1}{|G_{ER}|} log
\frac{1}{|G_{ER}|},\nonumber\\
 &=& log |G_{ER}| 
\end{eqnarray}
However it will not be necessary to compute $H(G)$ directly since we will only
be concerned with the change in entropy when random edges are inserted and
deleted.

Using $H(G)$ as the entropy of the graph $G$,
we can see that after each time step, the entropy of the graph has increased,
such that:
\begin{eqnarray*}
H(G_{i+1})&\ge&H(G_i)
\end{eqnarray*}
The proof is the following.  Define, $B$ as the graph before a given time
step, and $A$ as the graph after a given time step.
Using the notation $p(a)=Prob(A = a)$
The conditional entropy $H(A|B)$ is given from the conditional probability
distribution:
\begin{eqnarray*}
H(A|B) &=& \sum_b p(b) H(A|B = b)\\
       &=& - \sum_b p(b) \sum_a p(a|b) \log p(a|b)\\
       &=& - \sum_{a,b} p(a,b)  \log p(a|b)
\end{eqnarray*}
Using the mutual information\cite{CoverThomas} $I(A;B)$ we obtain:
\begin{eqnarray*}
I(A;B)&=&H(A)-H(A|B)\\
&=&H(B)-H(B|A)\\
H(A)-H(B)&=&H(A|B)-H(B|A) \ .
\end{eqnarray*}
Since $Prob(A = a | B = b)$ is given by the update rule,
where $a$ is the graph after an update, and $b$ is the graph before an update,  
$H(A|B)$ can be computed from the graph update rule.  There are $E_m$
maximum edges in the graph, and $E$ edges at a given time.
In a directed graph $E_m = N(N-1)$.
The entropy of our
random edge selection is $\log E$ which is the $log$ of the number of edges
from which the random selection is made.  We add an edge by selecting a random edge
to add, from all the edges that are not there.  The change in entropy of this
operation is $\log (E_m - (E-1))$ which is the $log$ of the number of absent
edges.  Thus:
\begin{eqnarray*}
H(A|B)&=&\log E + \log (E_m - (E-1))
\end{eqnarray*}
Computing $H(B|A)$ will in general depend on the prior distribution $P(B)$,
however we know that the prior can only \emph{reduce} entropy from the
maximum.  We don't know which edge was just added to $A$, but no matter what
$P(B)$ is, an upper bound on the entropy is the uniform assumption:
giving entropy $\log E$.
Likewise, if we know $P(B)$, we can find a probability distribution on which
edge was the edge which was deleted to arrive at $A$, however, the most the
entropy can be is $\log (E_m - (E-1))$, thus:
\begin{eqnarray*}
H(B|A)&\le& \log E + \log (E_m - (E-1))
\end{eqnarray*}
Which gives us the principle that entropy is never decreasing:
\begin{eqnarray*}
\Delta H&=&H(A)-H(B)\\
&=&H(A|B)-H(B|A)\\
&=&\log E + \log (E_m - (E-1)) - H(B|A)\\
&\ge& 0
\end{eqnarray*}

So, entropy can never decrease.  If the number of edges in a graph is fixed,
the maximum entropy distribution means that all graphs with that fixed number
of edges are equally likely.  If we have a large ER graph with $p=E/(N(N-1))$,
we expect there to be $E$ edges.  The Chernoff bound tells us that the
probability that an ER graph has more or less than $E$ edges falls
exponentially.
Thus, we expect the system to tend towards an ER random graph.
The binomial distribution means that each node will have a degree which is
close to the mean degree and that nodes with degree much higher or much lower
than average are rare.

Dorogovtsev et. al. \cite{sd2002statmech} discuss a similar 
dynamical model, where the total number of edges is fixed
and an edge rewiring scheme is introduced. 
However, the analysis is done with a view towards the production of 
power-law graphs and arbitrary fat-tailed distributions.

The model presented in this section
can be modified to one where only the expected number
of edges is constant, however the algebra will become more complex. 
As shown analytically in Section \ref{sec:rate_eq}
and via simulations in Section
\ref{sec:walks_simulations}, we can relax the
constraint that the number of edges be fixed,
make the numbers deleted and inserted random variables, and the result
will remain an ER graph. 

\section{Rate-Equations for the In-Degree Distribution}
\label{sec:rate_eq}
In this section we provide a rate-equation based analysis of the in-degree 
distribution of nodes in a stochastic network system similar to the one
introduced in section \ref{sec:maximal}. While the entropy analysis considered
graphs that lose and gain exactly one edge each time step, in the rate equation 
approach the number of edges created and destroyed are both random variables
that are chosen to produce a constant average in-degree.
In particular, let us consider the dynamics from the 
perspective of the nodes. For reasons that will be apparent in the section 
on load balancing, we will denote the average number of {\em absent edges} 
in the graph as $J$, and assign an integer $C$ as the 
{\em maximum in-degree} of any node; $C=N-1$ corresponds to the case 
considered in the previous section. Thus, the average number of edges
(which is the same as the sum of the in-degrees of all the nodes) in the 
network is $E=NC-J$. 

Let a randomly picked node have in-degree $i$ during a certain step.
Each edge is deleted uniformly and independently,
thus the probability that a node
with $i$ incoming edges will lose at least one edge is approximately
proportional to its in-degree.  Note that the expected number edges deleted 
per step in the whole network is $1$, and hence, we can ignore the case 
where more than one incoming edge is deleted at the same node. Hence, the 
rate at which the node's degree will decrease by one can be approximated as:
\begin{equation}
\mu_i=
\begin{cases}
\frac{i}{N C -J} & \text{$(0 \le i \leq C)$}\cr
0 & \text{else}\cr
\end{cases}
\label{eq_mu}
\end{equation}
Similarly, the rate at which it will acquire an edge (i.e., its degree will 
increase by one) can be assumed to be  proportional to   $(C-i)$ (i.e., the 
number of possible incoming edges that are absent):
\begin{equation}
\lambda_i= 
\begin{cases} 
\frac{C-i}{J} & \text{($0\leq i \le C$)}\cr  
0 & \text{else}.\cr
\end{cases}
\label{eq_lambda}
\end{equation}
The above process then can be seen as a birth and death
Markovian process with state-dependent arrival and service(departure) rates. 
Markov processes like this often appear in queueing theory and in that context this 
is an M/M/$\infty$//M queueing system\cite{kleinrockqueueing}.

In our situation, the states of the Markov process correspond to the 
instantaneous in-degree of a node (hence, the total number 
of states is $C+1$), and the rates at which its degree decreases and 
increases are given by Equs.~(\ref{eq_mu}), and (\ref{eq_lambda}), 
respectively. If the probability of being in state $n$ (i.e., the 
probability that a randomly picked node has degree $n$) is $p_n$, then the 
steady state distribution satisfies the following:
\begin{eqnarray*}
B &=& \{j : 1 \leq j < C \} ,\nonumber\\
0 &=& -(\lambda_j + \mu_j) p_j + \lambda_{j-1} p_{j-1} + \mu_{j+1}
p_{j+1}\nonumber\\
&& \qquad (\forall j \in B ).
\end{eqnarray*}
Which we rewrite as:
\begin{equation}
(\mu_{j+1}p_{j+1} -\lambda_j p_j) = (\mu_j p_j -\lambda_{j-1}p_{j-1}).
\label{eq_mc}
\end{equation}
The boundary condition is:
\begin{eqnarray*}
0 &=& -\lambda_0 p_0 + \mu_1 p_1.
\end{eqnarray*}
By solving equation \ref{eq_mc} we 
obtain a simple steady-state solution for the expected distribution
of jobs per node.
To solve this difference equation define
$\alpha_{j+1} = \mu_{j+1}p_{j+1} - \lambda_j p_j$, and note the equation
becomes:
\begin{eqnarray*}
\alpha_{j+1}&=&\alpha_j\\
0&=&\alpha_0
\end{eqnarray*}
Thus, $\alpha_j = 0$, or equivalently:
\begin{eqnarray*}
p_{j+1}&=&\frac{\lambda_j}{\mu_{j+1}}p_j.
\end{eqnarray*}
The solution of the above is:
\begin{eqnarray}
p_n &=& p_0 \prod_{i=0}^{n-1}\frac{\lambda_i}{\mu_{i+1}}, \nonumber\\
&=& p_0\frac{C (C-1)(C-2)\cdots (C-n+1) J^{-n} }{ n! (NC-J)^{-n}},\nonumber\\
&=& p_0 \left(\frac{NC-J}{J }\right)^n {C \choose n}.
\label{eq_ss}
\end{eqnarray}
We know that $\sum_n p_n = 1$, and 
thus we see that $p_n$ is binomially distributed:
\begin{eqnarray*}
p_n &=& {C \choose n}\left(1-\frac{J}{NC}\right)^n
\left(\frac{J}{NC}\right)^{C-n}.
\end{eqnarray*}
If we define a normalized quantity as
$\alpha=J/NC$ ($\alpha$ will have a physical meaning in the context of the 
load balancing system discussed in section \ref{sec:load_balancing} ), then we get the variance of 
the distribution to be
$\sigma^2=C\alpha(1-\alpha)$.

If $C=N-1$, then we see that this model recovers
the ER graph we found in Section \ref{sec:maximal}:
\begin{eqnarray*}
p_n &=& {N-1 \choose n}q^n (1-q)^{N-1-n}.
\label{eq_re_sol}
\end{eqnarray*}
Where $q=\frac{E}{N(N-1)}$ and $E=N(N-1)-J$.

\section{Local Dynamics and Simulation Results}
\label{sec:walks_simulations}
Implementing the exact graph dynamics discussed in the two previous
sections would require global knowledge about individual node's degrees
and/or the number of edges in the network at any step. However, it would be
ideal to be able to have a stochastic dynamic that just samples the graph
using local information, and makes decisions about which edges to add or
delete.  Eq.~(\ref{eq_mu}) provides a useful  lead: it implies that a node 
loses an edge {\em preferentially} with respect to its in-degree.
We know that if one performs a long-enough random walk
on an undirected graph (or a Eulerian directed graph)
then in steady state the 
probability that the walk will end at a particular node is proportional to 
its in-degree\cite{lovasz95mixing}.
The issue is about the required length of the random walk. 
Ideally we would like to do a walk of length  no more than $O(\log N)$; we 
justify in the following remarks why a logarithmic length random walk 
suffices for our case and also provide simulation evidence.

\begin{figure*}
\begin{center}
\includegraphics[scale=1.0,angle=0]{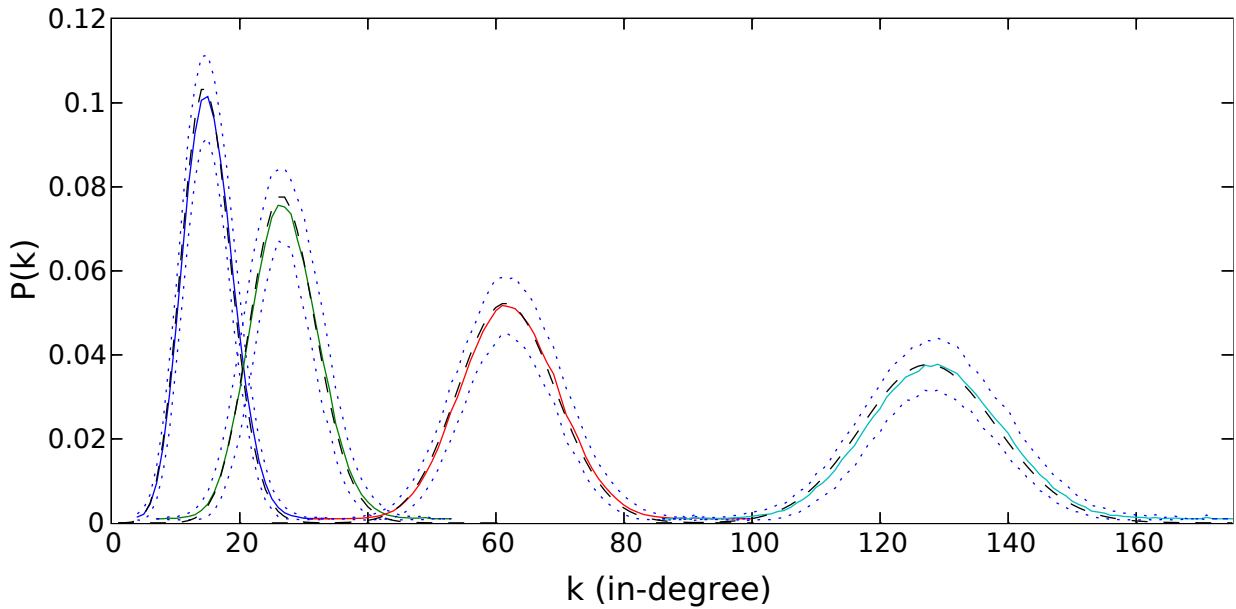}
\caption{The proposed edge dynamic generates Erd\"{o}s-R\'{e}nyi random graphs.
The steady state in-degree distributions are shown for the
random-walk based network dynamics described in section 
\ref{sec:walks_simulations}.  These graphs have $N=1024$, maximum 
in-degree $N-1$, and average in-degrees (from left to right) of $16$, 
$32$, $64$ and $128$ respectively.  Each graph begins as a
completely structured graph with $O(N)$ diameter. After the 
edge dynamics thermalizes the graph, the result is an Erd\"{o}s-R\'{e}nyi random graph with 
$O(\log{N})$ diameter. The mean distributions 
(solid lines) are each shown with upper and lower standard deviation range 
curves (dotted lines). The predicted binomial degree distributions 
(dashed lines) are shown for comparison. Arrivals and 
departures at each time step are Poisson distributed.}
\label{fig:in_degree_unrestricted}
\end{center}
\end{figure*}

In order to verify that local dynamics do indeed lead to ER graphs and 
match the theoretical predictions, we performed extensive simulations with 
several protocols and the results are reported in this and the following 
sections. We intentionally introduced certain deviations from the exact 
protocol discussed in sections \ref{sec:maximal} and \ref{sec:rate_eq} so as 
to demonstrate the robustness of the whole system and keep the protocols 
relevant for load balancing systems studied in this paper.  

\begin{figure}
\begin{center}
\includegraphics[scale=1.0,angle=0]{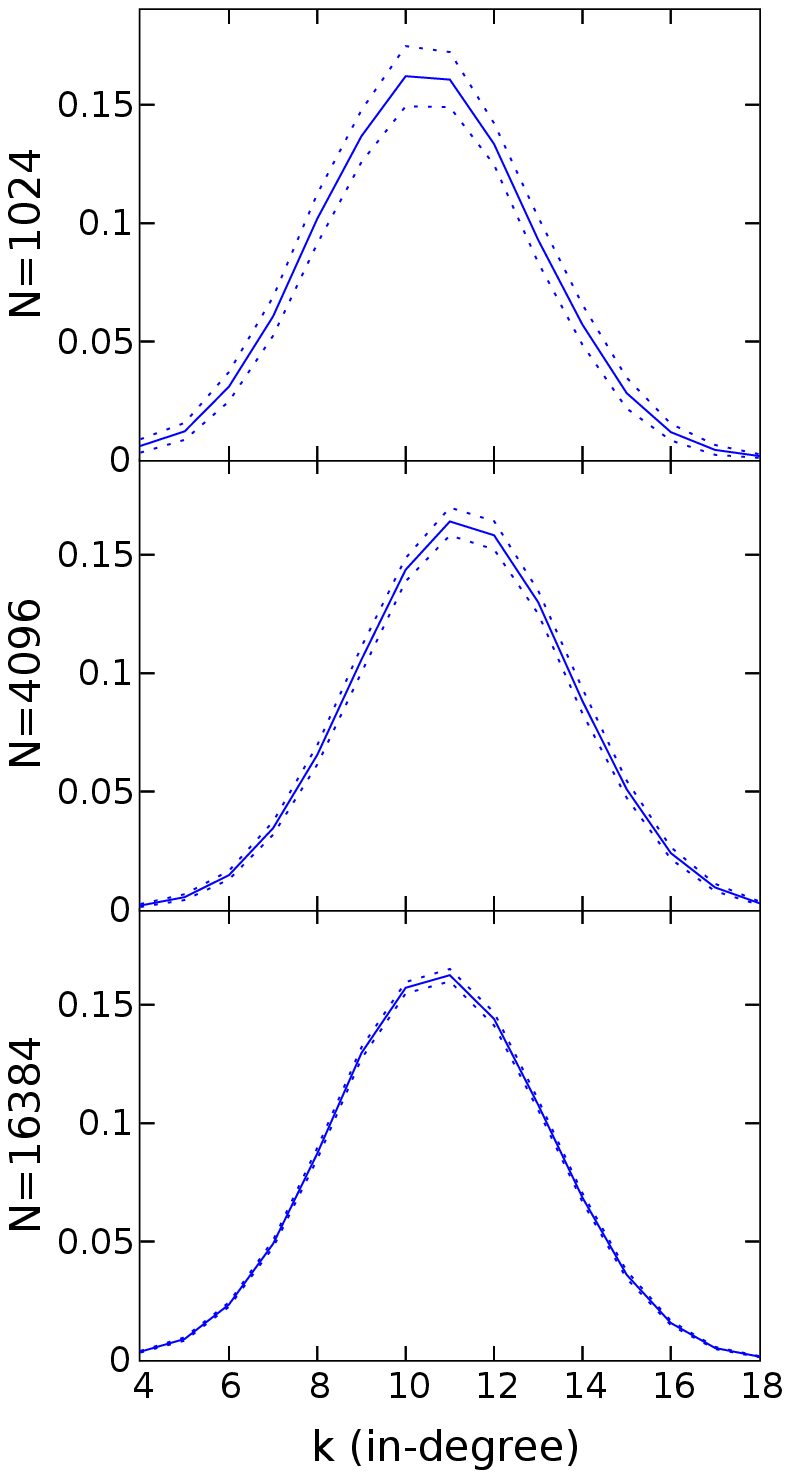}
\caption{The production of Erd\"{o}s-R\'{e}nyi random graphs using the random-walk based
dynamics discussed in section \ref{sec:walks_simulations} is effective for 
all network sizes simulated (at least up to $N=16384$). These in-degree 
distributions (solid lines) are for graphs with $N=1024$, $N=4096$ and 
$N=16384$. Standard deviation range curves are shown as dotted lines.  
These graphs have an imposed maximum degree, $C= 22$, and a mean in-degree 
of approximately $11$. With increasing $N$ we see tighter convergence to 
the binomial distribution.  Arrivals and departures at each time
step are Poisson distributed.}
\label{fig:size_scaling_in_degree}
\end{center}
\end{figure}

\begin{figure}
\begin{center}
\includegraphics[scale=1.0,angle=0]{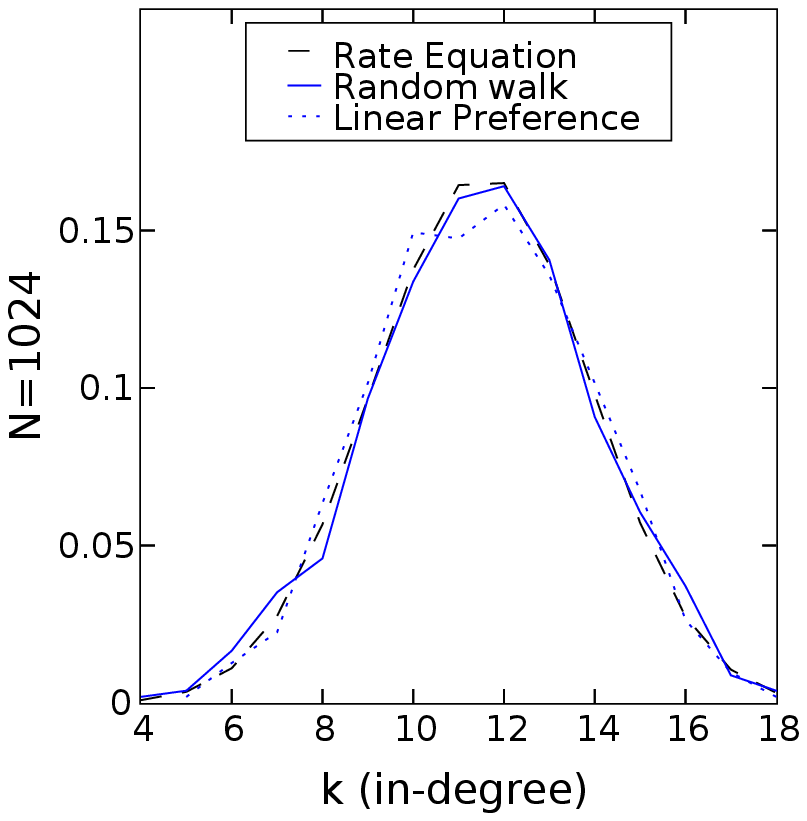}
\caption{The simulated networks generated by short random walk dynamics with a fixed
number of edges are very similar to networks produced by global linear preferential 
selection dynamics.  The random walk job migration (solid line) and the
global preferential job allocation (dotted line) described in section 
\ref{sec:rate_eq} Eq.(\ref{eq_mu}) produce in-degree distributions that match
well to the predicted binomial distribution. The random walk simulation graph 
has a constant number of edges and is thus very similar to the rewiring dynamic 
described in section \ref{sec:maximal} . The linear preference simulation uses 
global linear preference to distribute jobs rather than 
the local random walk approximation. Along with Fig.\ref{fig:mixing_correlation}
this shows that random walks can be used to approximate linear preferential
selection of nodes. Single snapshots of the distributions are depicted rather than 
distribution averages shown in Figs. \ref{fig:in_degree_unrestricted}, 
\ref{fig:size_scaling_in_degree} and \ref{fig:out_degree_unrestricted}. 
Table \ref{tab:sim_results} summarizes the simulation parameters and distribution 
variances.} 
\label{fig:in_degree_restricted}
\end{center}
\end{figure}

\noindent
{\bf 1.} {\em Graph Initialization:} First we create a directed graph  with 
$N$ nodes and $E=N\ave{k}$ edges  such that the maximum degree of any node 
is $\leq C$ (note the case of no restrictions on the maximum degree, i.e., 
$C=N-1$, is considered in Figs. \ref{fig:in_degree_unrestricted} where as 
the case of restricted maximum degree is considered in 
Figs. \ref{fig:size_scaling_in_degree} and \ref{fig:in_degree_restricted} ). 
This graph is intentionally constructed in 
a very structured fashion so as to show that the proposed dynamics do 
indeed lead to a random graph independent of the initial configuration. The
initial structure is created by connecting node $i$ by incoming edges to 
nodes $(i+1) \mod N$,  $(i+2) \mod N$, $\cdots$, and $(i+\ave{k}) \mod N$, where 
$\ave{k}$ is the average degree of the nodes. Further evidence that the dynamics
effectively randomize the initial graph can be seen from the fact that the 
initial graphs have a diameter of $O(N/\ave{k})$ while the thermalized graphs 
have diameter bounded above by $O(\log N)$.

\noindent
{\bf 2.}{\em Edge deletion:} For this set of simulations, at each time step 
we delete a Poisson-distributed number of edges. A random walk of length $\log N$ 
(i.e., of lengths 10 ,12 and 14 in the simulations reported in Figs. 
\ref{fig:size_scaling_in_degree} ) is initiated 
from a fixed particular node, and the last node on the random walk randomly deletes 
one of its incoming edges.

\noindent 
{\bf 3.} {\em Edge Insertion:} Ideally, we should insert an edge picked 
randomly among the absent edges. One way to achieve this would be to do a 
random walk on the complementary graph $\bar{G}$ (i.e., where the absent 
edges in $G$ are present) and pick the node it ends at as the arrow-end of 
the edge to be inserted; i.e., the node, $i$, the random walk ends at gets 
an incoming edge.  To decide on the node that will get the outgoing edge, 
one could reverse the directions of the edges in the complementary graph 
$\bar{G}$, and perform a random walk starting at the node $i$ (found in the 
previous step), and select the node,  $j$, that the random walk ends at. 
Then a directed edge from node $j$ to node $i$ is added. The two steps ensure 
that both the in-degree and the out-degree of a node increase proportional 
to its respective absent degree (i.e., the number of all possible edges that 
are missing; see Eq.~(\ref{eq_lambda})). 

\begin{figure*}
\begin{center}
\includegraphics[scale=1.0,angle=0]{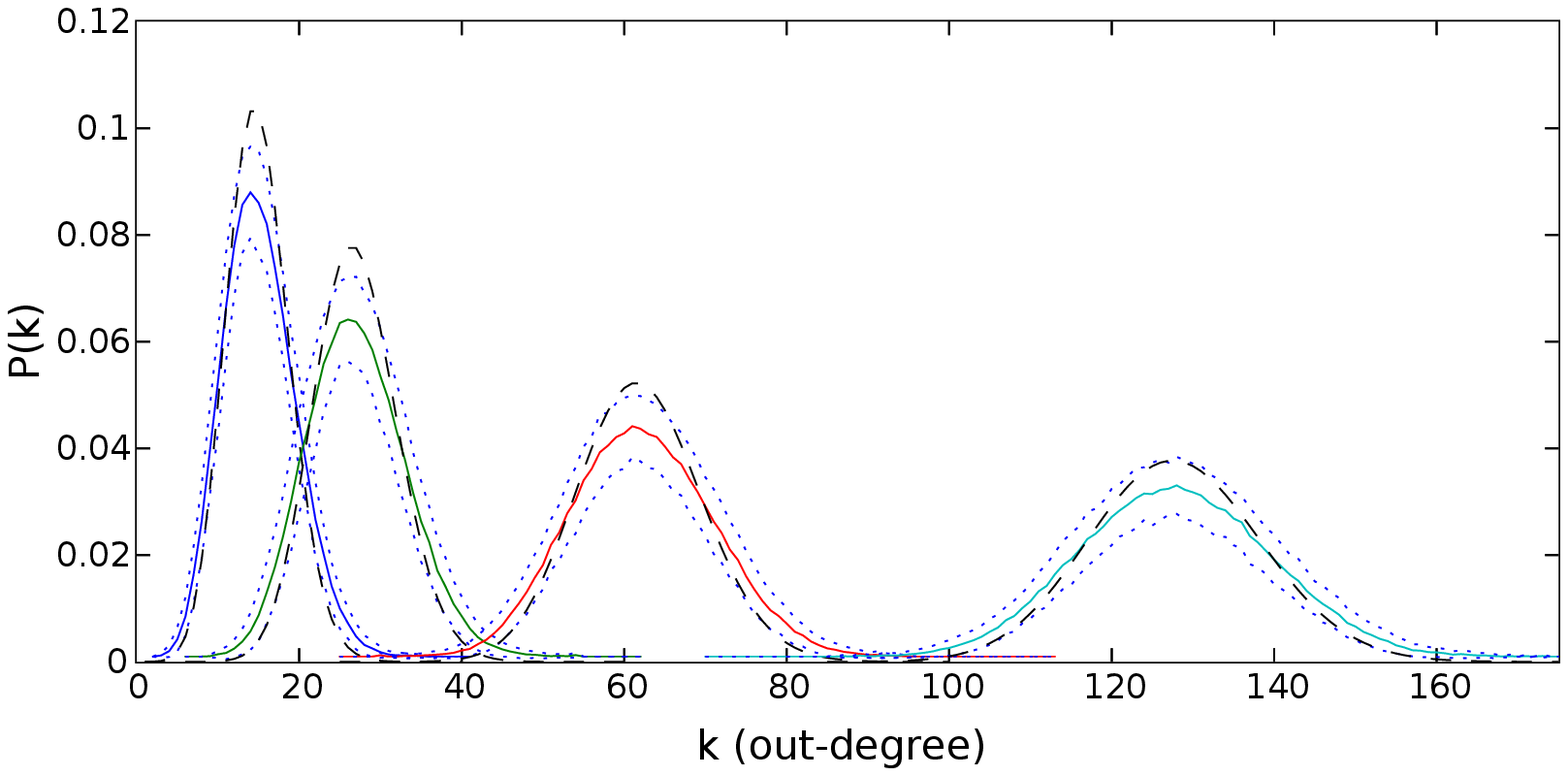}
\caption{ We see that these simulated graphs 
have out-degree distributions that deviate slightly from the binomial degree 
distributions. These out-degree distributions are from the same simulations 
presented in Fig.\ref{fig:in_degree_unrestricted}. These graphs have $N=1024$, 
maximum in-degree $N-1$, and average in-degrees (from left to right) of $16$, 
$32$, $64$ and $128$ respectively. The out-degree distributions have similar 
averages but larger variances than the in-degree distributions. The mean 
out-degree distributions (solid lines) are each shown with upper and lower 
standard deviation range curves (dotted lines). Given that the in-degrees and 
out-degrees are subject to different dynamics we would not expect the out-degree 
distribution to follow the binomial distribution. Arrivals and departures at each 
time step are Poisson distributed.}
\label{fig:out_degree_unrestricted}
\end{center}
\end{figure*}

In our protocol, however, we take into consideration the fact that a random
walk on the complementary graph is not needed or feasible in certain
applications. Moreover, in the load-balancing application considered in the
next section the absent edges of the graph correspond to jobs that are
occupying the resources of the node. Therefore, the addition of an edge is 
undertaken when a node finishes one of its jobs and wants to increase its 
in-degree. Thus, in one simulation reported in
Fig.\ref{fig:in_degree_restricted} and Table \ref{tab:sim_results}, 
we assume that a node is chosen randomly proportional to its missing degree 
to receive an incoming edge. Since missing degrees are proportional to the number 
of jobs on a node, the above assumption is equivalent to assuming that each job ends 
with equal probability at each time step (or that job length follows a geometric 
distribution). This decision is made globally; the situation where each node 
independently makes its decision to accept an incoming edge (thus following 
Eq.~(\ref{eq_lambda}) exactly) based on the detailed simulated execution of 
resource-consuming jobs on resource-bearing nodes will be reported in future 
work. Now in order to find the other end of the edge, instead of performing a 
random walk on the complementary graph (with edges reversed), we still perform 
a random walk on the graph $G$ starting at the node selected to 
receive an incoming edge, and select the node that the random walk ends at.  
This has the consequence that the out-degree distribution will deviate 
somewhat from the ER prediction (see Fig.\ref{fig:out_degree_unrestricted}). 
For the purpose of the proposed load balancing system this is not a 
liability since the out-degree of a node is not physically meaningful.

Here are a few additional remarks about the graphs generated and our 
simulation results and local random-walk based protocols:

\noindent
{\bf 1.} {\em Minimum Degree of a node}: The connectivity and diameter of 
random graphs are both well-established and are critical measures of 
performance.  For instance, in order for a random graph to have a giant 
component, the average degree, $\langle k \rangle$, must be greater than 2. 
If $\langle k\rangle > 7/2$, then the diameter of a random graph scales 
logarithmically\cite{albert-statistical}:

\begin{equation}
d(k,N) \propto \frac{\ln N}{\ln\langle k\rangle}
\label{EQ_ER_diameter}
\end{equation}
These results apply to undirected random graphs and since we are focused on
directed graphs we provide supporting simulation results to show that this 
protocol produces connected, strongly-connected and fast mixing directed 
graphs with low directed diameter.

If the network does not have a giant component, then many nodes will be 
isolated and unable to participate in a local search algorithm.  The 
implications for load balancing applications will be discussed in section 
\ref{sec:walks_simulations}.

\begin{figure*}
\begin{center}
\includegraphics[scale=1.0,angle=0]{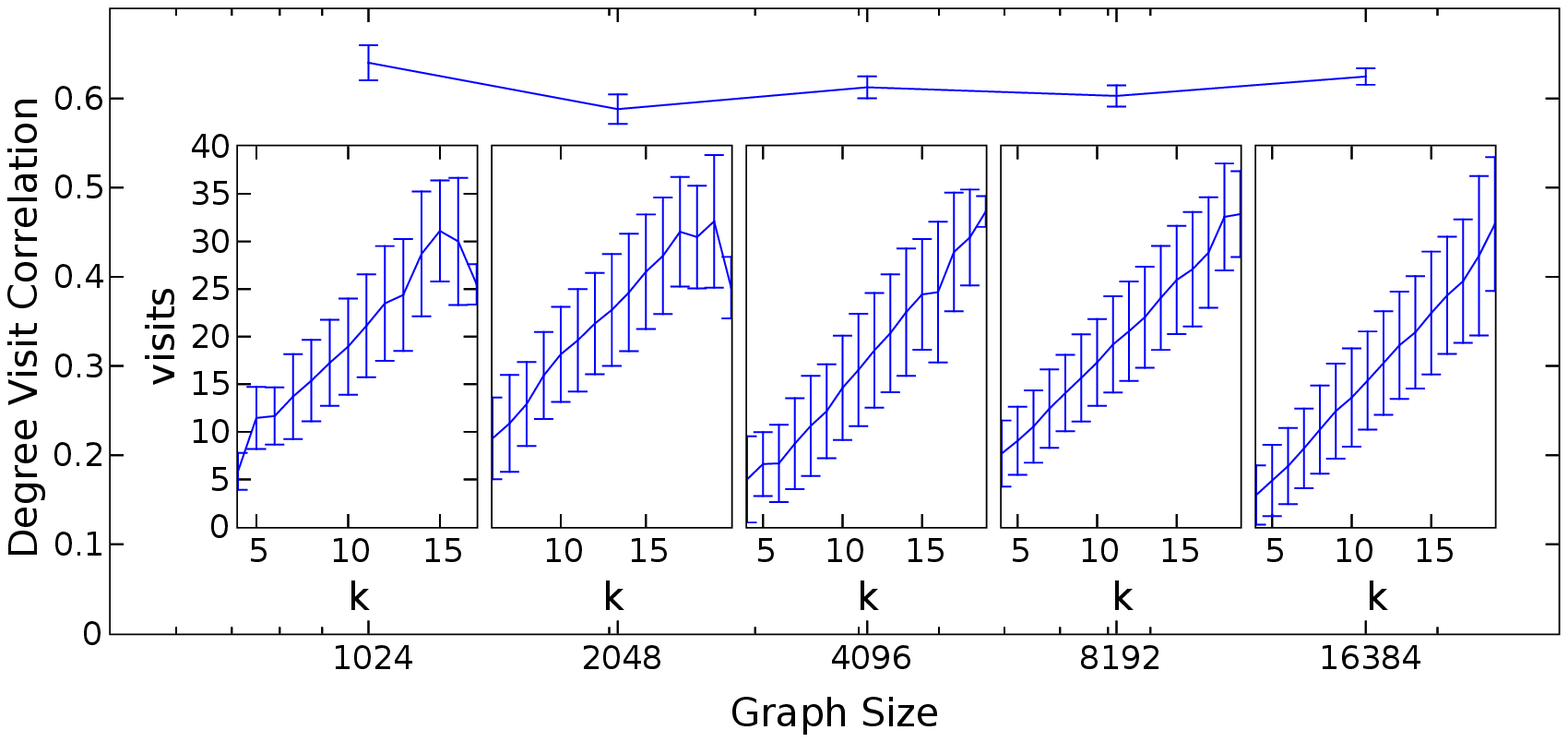}
\caption{ Short random walks are a good approximation to linear preferential
selection. In Fig. \ref{fig:in_degree_restricted} we observed
that random walk dynamics and linear preferential selection produced similar 
degree distributions.  Here we show directly that each node's in-degree, $k$, 
is correlated  to the frequency that random walks terminate at that node; 
we also demonstrate that this relationship holds for a range of network sizes. 
The correlation coefficient averaged over 100 snapshots of the graph is shown 
(solid line) at the top of the plot with standard deviation range bars. The 
consistent average correlation indicates that the approximate linear preference 
observed for short random walks scales over a range of networks sizes that 
exceeds an order of magnitude($N=1024$ to $N=16384$). The insets show that nodes 
have a number of visits linearly proportional to $k$ . In each inset the mean 
number of visits (solid line) and standard deviation range bars are plotted vs.
node in-degree. Except for the minor deviations seen for $N=1024$ and
$N=2048$, we see that short random walks produce a very close approximation to
linear preference. In all cases the length of each random walk is $\log N$.}
\label{fig:mixing_correlation}
\end{center}
\end{figure*}
Ideally we want the graph to have no disconnected components and be 
strongly-connected; this will ensure that our random walk approach can properly
thermalize the graph. Strictly speaking, we do not need the graph to have a
single strongly-connected component at every step since the introduction of 
new edges should repair the graph and make it a single strongly-connected 
component often enough. The repairs can happen because the direction of an
edge only affects the propagation direction for the random walks.  An example
of a connected (but not strongly-connected) graph is a pair of 
strongly-connected clusters, $A$ and $B$, with a single edge $e_{AB}$ going from a
node in $A$ to a node in $B$. Random walks initiated in $B$ will never reach
nodes in $A$ unless an edge from $B$ to $A$ is created.  But such edges can be
easily formed.  When a node in $A$ needs a new incoming edge it initiates a
walk to find a node to be the other side of the edge.  When such a walk
crosses $e_{AB}$ and ends at a node in $B$, a new edge $e_{BA}$ will be
created that goes from $B$ to $A$.  

In our simulations we do indeed find that strongly-connected components
split and merge as the graph evolves.  However in all simulations conducted
with an imposed minimum degree of $4$ we observe that the network never
permanently fragments into multiple strongly-connected components. 
Fluctuations occur but the graphs heal themselves. Although this does not 
constitute a proof that the graphs will always remain strongly-connected, we 
observe in all simulations that the number of strongly-connected components remains 
near $1$. This is an important practical concern for an implementation and more 
detailed simulation and analysis will be the subject of future work.

\noindent
{\bf 2.} {\em Mixing time and the length of the Random Walks}: It is known 
that for a random $d$-regular undirected graph, the mixing time (i.e., the length 
of the random walk necessary to sample edges uniformly, or equivalently, 
nodes preferentially) scales as $O(\log N)$, where $N$ is the number of 
nodes in the network\cite{kahake-expander}. It is widely believed that since 
an ER graph is almost 
a $\ave{k}$-regular graph, where $\ave{k}$ is the average degree of the nodes, the 
mixing time should also scale as $O(\log N)$.   If the average degree is 
$\log N$, then one can prove this result; however, a formal proof is not 
there for ER graphs with constant average degree. For our case, however, 
$\log N$ average degree is quite reasonable; if $N=1048=2^{10}$ then the 
average degree has to be $10$ for the formal results to hold. 

A related quantity to mixing time is the graph diameter.  In order to sample
edges uniformly using a $O(\log N)$ random walk the diameter of the graph
cannot be larger than $O(\log N)$ since the walk must be able to reach every
edge to sample uniformly. Starting with a structured graph with $O(N)$ diameter 
the thermalized graph that results from a few thousand iterations now has a 
diameter that is of $O(\log N)$ in all simulations conducted.

In order to verify that the random walks do indeed sample the nodes
proportionally to their in-degree in these directed networks, we provide the
following simulation results. After the evolving graph structure has stabilized we
freeze the graph and select a node that initiates $20 N$ random walks of
length $(\log N)$. We record the number of times that each node is the last
visited on a walk and then calculate the correlation between each node's 
visitation frequency and in-degree. Figure \ref{fig:mixing_correlation} shows 
a high correlation coefficient.  Also note that each node is visited at least 
once in all simulations which confirm the low diameter of the graph.  Each node 
in the starting graph was selected to initiate the walks to show that the 
mixing is uniform throughout the graph.

{\bf 3.} {\em Out-Degree Distributions:} Although the in-degree distribution
is  of primary interest in this paper, we also briefly report the out-degree
distribution here. In Figure \ref{fig:out_degree_unrestricted} we see that 
the out-degree distribution does not match the ER distribution as closely as 
the in-degree. This lack of symmetry is due to the fact that the random walks 
performed when adding and removing edges travel over directed connections. If 
we wished the out-degree distribution to follow an ER degree distribution as 
closely as the in-degree we could follow a similar protocol for both for both
in-degree and out-degree. However in this protocol the out-degree is not of
interest and will not be considered in detail..

\begin{table}
\begin{tabular}{|c|c|}
	\hline
System type &  $\sigma_{free}^2$    \\
	\hline
Rate equation prediction   &   5.49\\
Linear preference simulation & 5.59 \\
Random walk simulation & 5.81 \\
	\hline
\end{tabular}

\caption{
Variances for random walk and linear preference simulations shown in Fig.
\ref{fig:in_degree_restricted} as well as the
Binomial distribution maximum degree $22$,
$1024$ nodes, and $10752$ jobs.    
}
\label{tab:sim_results}
\end{table}

%\begin{table}
%\begin{tabular}{|c|c|c|}
%	\hline
%N &  $\sigma_{free}^2$ & Correlation Coefficient    \\
%	\hline
%1024 & & \\
%2048 & & \\
%4096 & & \\
%8192 & & \\
%16384 & & \\
%\hline
%\end{tabular}
%
%\caption{
%Variances and mixing correlation information for simulations with 
%maximum degree $22$ and mean targeted degree $11$.  The simulation is repeated
%for a range of network sizes and the walk lengths are $\log(N)$. The large
%correlation coefficients indicate that the random walks effectively mix and
%sample from the underlying degree distribution.
%}
%\label{tab:scaling_sim_results}
%\end{table}

\section{A Load-Balancing Paradigm}
\label{sec:load_balancing}
\subsection{Previous work}
The field of load-balancing has been active for decades and many techniques
and problem formulations have been used to approach the problem
\cite{lling91study, 1988casa, 1996luis, mm2001twochoice}. The use of random 
walks has produced some interesting empirical load balancing results in sensor 
networks \cite{ca2004randomwalk}. What distinguishes the proposed protocol
from prior work is the use of random walk sampling on an overlay network whose 
topology is actively shaped by the dynamics of the protocol. No monitoring is
performed in this scheme since the load balancing algorithm and state
information is encoded in the overlay network structure. Please note that this 
overlay graph need not consist of physical links as long 
as the network is globally addressable. Since the Internet is addressable, each 
node in an Internet-based overlay will only need to maintain a table of of its 
neighbor nodes rather than a physical connection for each neighbor. On the other 
hand in networks that are not globally addressable the overlay edges will need 
to contain complete route information or the edges will need to be actual 
physical links.

\subsection{Statistical Mechanical Load-Balancer}
Let us take the same statistical mechanical system as in the preceding 
sections and encode it as follows: \\
(i) Each node represents a server or processor providing service to a 
networked community. \\
(ii) The in-degree of a node represents the amount of {\em free resources} 
of the particular node, e.g., the number of extra jobs in can handle.\\
(iii) The maximum in-degree, $C$, is the {\em maximum capacity} that each 
node in the network can handle.\\
(iv) The state of the network, i.e., the in-degree distribution, represents 
how {\em balanced} the load distribution is.\\

In the steady state, jobs/requests arrive at the same rate as the jobs are 
completed by the suite of servers. Hence, when a job arrives (by random walk), 
in order to represent the node's increased load and decreased free resources, 
we may need to decrease the in-degree of the node that received the new job; 
this is done by deleting one of its incoming edges uniformly randomly. Similarly, 
when a job completes, the corresponding server may have excess free resources 
and it should indicate its new state by increasing its in-degree; this is done 
by performing a short random walk and making a new directed edge that originates
at the last node on the random walk and terminates at the node that initiated 
the walk.

The outline of the steps is summarized below.
\begin{itemize}
  \item Create a graph $G$ whose nodes have in-degree proportional 
  to free resources.
  \item When a node, $v_{i}$, creates new load, it performs a
  short random-walk on $G$ and distributes the new load to the end node 
  on the walk.
  \item Nodes compensate for changes in load by creating or deleting edges
  in accordance with the prescribed edge dynamic which keeps the in-degree of
  a node proportional to its free resources. Newly arrived jobs can delete edges 
  while jobs that are completing can create new edges.
\end{itemize}

The dynamics we have proposed in the previous section create and maintain the
structure of the overlay network through the deletion and creation of edges.
This resulting overlay network is an ER graph (or its variant studied in 
section~\ref{sec:rate_eq} that has a bounded number of in-degrees $C$) which is 
an almost regular graph. 

Mapping the properties of this overlay network 
to node resources can be handled easily by making the natural assumption that there 
is a scalar metric, $R$, that each node can locally calculate to determine its free 
resources. For web mirroring the relevant metric is the bandwidth that the next request 
can expect to receive.  It would be calculated as the peak outgoing 
bandwidth $B_{i}^{(max)}$ divided by the current number of requests $D_{i}$ plus 
$1$: $R_i=\frac{ B_{i}^{(max)} }{D_i + 1}$.  The nodes agree on a size for the 
unit of capacity represented by an in-degree.  Using this unit of capacity, each
node maintains a targeted in-degree proportional to free resources:
\begin{eqnarray}
\label{eq:ki_of_r}
k_i(t) &=& \max(\frac{C R_i }{B_{i}^{(max)}}, k^{(min)}) ,\nonumber\\
       &=& \max(\frac{C}{D_i +1}, k^{(min)})
\end{eqnarray}
Distributed mechanisms to alter the mapping of resources to in-degree can be added to 
account for changes in capacity and network size but we will not consider such details 
here. 

\section{Discussion}
We present a distributed algorithm that generates Erd\"{o}s-R\'{e}nyi random 
graphs.  Rate equation calculations and maximum entropy arguments predict that
this protocol will yield Erd\"{o}s-R\'{e}nyi graphs and simulation results spanning a large 
range of sizes and average degrees support that prediction. The agreement of 
the simulations and the predicted Erd\"{o}s-R\'{e}nyi degree distribution is excellent and the 
diameters of the resulting networks are $O(\log N)$ as we would expect for a 
random network.  Short random walk sampling  shows that there is a high 
correlation coefficient between a node's degree and the frequency that a 
random walk terminates at that node which justifies the use of short random 
walks as a decentralized substitute for linear preferential attachment.

These emergent Erd\"{o}s-R\'{e}nyi graphs can be used to provide a scalable resource allocation 
platform that does not rely on any central authority to distribute load.  All 
operations are local and the latency for resource discovery (random walk)
is $O(\log N)$.  The most obvious applications are for WWW mirroring and 
distributed computing\cite{dm2000migraiton},  but the same idea is applicable 
whenever there is a large set of servers that provide the same service and users 
who want jobs to be done.

In the case of grid computing, we can imagine a large set of nodes connected
according to our algorithm.  When one node is overwhelmed by work, it can 
make use of unused computing power in the grid.  In the case of
non-communication-bound jobs (such as various optimization problems), clearly
this system will work well.  More work is needed to study the applicability 
to the general case of distributed computing, namely where the jobs
are rather short and depend on the output of many other jobs. The types of
communications-bound and distance-sensitive situations will be treated in
future work. 

Our algorithm is also ideal for content mirrors on the Internet.  Each server 
might have a certain amount of bandwidth which is sliced into quantized 
units. Each server has an incoming edge for each unit of bandwidth it can 
offer. When and HTTP request comes to the server, a background process can 
migrate that request using a short random walk. Finally, using standard 
HTTP redirection codes, the client is redirected to the new server which 
allocates a unit of bandwidth to the client.

The main idea is to correlate resources with in-degree and then allow the
graph to thermalize under and edge dynamic.  Erd\"{o}s-R\'{e}nyi graphs have peaked binomial
distributions that decay exponentially and thus provide good load balancing.
However the ideal degree distribution in this case is a regular graph 
since every node has the same in-degree and thus the same load. Modifications 
to the random walk protocol to generate a more regular graph can further improve 
performance. We have developed such a protocol and its analysis and 
detailed simulation will be a topic for future work.

\bibliography{templ.bib} 
%make bibliography with bibtex

\end{document}